\documentclass[draft]{agujournal2019}
\usepackage{url} %this package should fix any errors with URLs in refs.
\usepackage{lineno}
\usepackage[finalnew]{trackchanges} %for better track changes. finalnew option will compile document with changes incorporated.
\usepackage{soul}

\newcommand{\corr}[1]{\textcolor{black}{#1}} %corrected parts
\draftfalse

\journalname{JGR: Earth Surface}

\begin{document}

%% ------------------------------------------------------------------------ %%
%  Title
%
% (A title should be specific, informative, and brief. Use
% abbreviations only if they are defined in the abstract. Titles that
% start with general keywords then specific terms are optimized in
% searches)
%
%% ------------------------------------------------------------------------ %%

\title{Barchans interacting with dune-size obstacles: details of the fluid flow and motion of grains}

%% ------------------------------------------------------------------------ %%
%
%  AUTHORS AND AFFILIATIONS
%
%% ------------------------------------------------------------------------ %%

% Authors are individuals who have significantly contributed to the
% research and preparation of the article. Group authors are allowed, if
% each author in the group is separately identified in an appendix.)

% List authors by first name or initial followed by last name and
% separated by commas. Use \affil{} to number affiliations, and
% \thanks{} for author notes.
% Additional author notes should be indicated with \thanks{} (for
% example, for current addresses).

% Example: \authors{A. B. Author\affil{1}\thanks{Current address, Antartica}, B. C. Author\affil{2,3}, and D. E.
% Author\affil{3,4}\thanks{Also funded by Monsanto.}}

\authors{N. C. Lima\affil{1}, W. R. Assis\affil{2}, D. S. Borges\affil{1}, E. M. Franklin\affil{1}}

% \affiliation{1}{First Affiliation}
% \affiliation{2}{Second Affiliation}
% \affiliation{3}{Third Affiliation}
% \affiliation{4}{Fourth Affiliation}

\affiliation{1}{Faculdade de Engenharia Mec\^anica, Universidade Estadual de Campinas (UNICAMP),\\
Rua Mendeleyev, 200, Campinas, SP, Brazil}

\affiliation{2}{Saint Anthony Falls Laboratory, University of Minnesota,\\
2 3rd Ave SE, Minneapolis, Minnesota, USA}
%(repeat as many times as is necessary)

%% Corresponding Author:
% Corresponding author mailing address and e-mail address:

% (include name and email addresses of the corresponding author.  More
% than one corresponding author is allowed in this LaTeX file and for
% publication; but only one corresponding author is allowed in our
% editorial system.)

% Example: \correspondingauthor{First and Last Name}{email@address.edu}

\correspondingauthor{Willian R. Assis}{righiassis@gmail.com}

%% Keypoints, final entry on title page.

%  List up to three key points (at least one is required)
%  Key Points summarize the main points and conclusions of the article
%  Each must be 100 characters or less with no special characters or punctuation and must be complete sentences

% Example:
% \begin{keypoints}
% \item	List up to three key points (at least one is required)
% \item	Key Points summarize the main points and conclusions of the article
% \item	Each must be 100 characters or less with no special characters or punctuation and must be complete sentences
% \end{keypoints}

\begin{keypoints}
\item Numerical results show that the disturbed fluid flow acts directly on the different types of subaqueous dune-obstacle interactions
\item The strength of a horseshoe vortex existing upstream the obstacle explains why grains either circumvent or pass over it
\item The distribution of the resultant force acting on each grain correlates well with the fluid trajectories of the disturbed flow
\end{keypoints}

%% ------------------------------------------------------------------------ %%
%
%  ABSTRACT and PLAIN LANGUAGE SUMMARY
%
% A good Abstract will begin with a short description of the problem
% being addressed, briefly describe the new data or analyses, then
% briefly states the main conclusion(s) and how they are supported and
% uncertainties.

% The Plain Language Summary should be written for a broad audience,
% including journalists and the science-interested public, that will not have 
% a background in your field.
%
% A Plain Language Summary is required in GRL, JGR: Planets, JGR: Biogeosciences,
% JGR: Oceans, G-Cubed, Reviews of Geophysics, and JAMES.
% see http://sharingscience.agu.org/creating-plain-language-summary/)
%
%% ------------------------------------------------------------------------ %%

%% \begin{abstract} starts the second page

\begin{abstract}
We investigate details of the interaction of subaqueous barchans with dune-size obstacles by carrying out numerical simulations where the fluid is solved at the grain scale and the motions of individual grains are computed at all time steps. With the outputs, we analyze the disturbances of the fluid flow, the trajectories of grains, and the resultant force on each grain, the latter being unfeasible from experiments and field measurements. We show that in some cases particles pass over the obstacle, while in others they completely circumvent it (without touching it), or are even blocked. For the circumvention and blocking cases, which we call bypass and trapped, respectively, we show the existence of a strong vortex between the lee face of the dune and the obstacle. This vortex results from the interactions of recirculation regions and horseshoe vortices, and has enough strength to deviate the main flow and carry grains around the obstacle in those cases. Our results shed light on the reasons for passing over, circumventing, and blocking, and contribute to our understanding of dunes in the presence of large obstacles such as hills, crater rims, and human constructions.
\end{abstract}

\section*{Plain Language Summary}
Barchans are crescent-shaped dunes, with horns pointing downstream, that are found on Earth, Mars and other celestial bodies. Although usually present on large extensions of desertic areas, barchans sometimes approach obstacles of comparable size, such as crater rims on Mars, houses and buildings on Earth, and bridge pillars when dunes are under water. In this paper, we investigate details of the interaction of barchans with dune-size obstacles by carrying out numerical simulations where the flow is solved with a relatively high spatial resolution (close to the diameter of grains) and the motion of grains is computed at all time steps. For all simulated cases, we show the trajectories of grains and the resultant force acting on each individual grain, the latter unfeasible to be obtained from field measurements. We show that in some cases particles pass over the obstacle, while in others they completely circumvent it (without touching it), or are even blocked, the flow disturbances being in the origin of these behaviors. Our results are important for better understanding how dunes behave in the presence of large obstacles, such as those found on Earth and Mars.

\section{Introduction}

Barchans are crescent-shaped dunes that usually grow by the action of a roughly unidirectional flow over a non-erodible ground with limited sand availability \cite{Bagnold_1, Herrmann_Sauermann, Hersen_3, Courrech}. These dunes are found on Earth (both in eolian and aquatic environments), Mars, and other celestial bodies, sharing roughly the same morphology, but having different scales \cite{Hersen_1, Elbelrhiti, Claudin_Andreotti, Parteli2}: up to one kilometer and millennium for Martian barchans, hundreds of meters and years for barchans on Earth's deserts, and centimeters and minutes for subaqueous barchans. However, despite the barchan shape being a strong attractor, variations of its morphology with hilly terrains \cite{Finkel, Bourke, Parteli4}, wind changes \cite{Finkel, Bourke, Parteli4, Assis5}, barchan-barchan collisions \cite{Long, Hersen_5, Bourke, Vermeesch, Parteli4, Assis}, polidispersity \cite{Alvarez6, Assis3}, and interactions with dune-size obstacles \cite{Assis4} have been reported.

Though usually present on large extensions of desert areas and subaqueous environment (with respect to the dune size), dunes sometimes approach obstacles of comparable size, such as crater rims on Mars \cite{Breed, Urso, Roback}, houses and buildings on Earth \cite{Raffaele}, and bridge pillars in subaqueous environment \cite{Rubi, Jia}. On Earth, dunes sometimes reach grasslands and mangroves, threatening biodiversity \cite<such as in Len\c{c}\'ois Maranhenses, Brazil,>{Amaral}, a situation that can be aggravated by climate change \cite{Baas}. On the other hand, sand dunes are also a natural protection for coastal regions, but human constructions are, however, wiping them out, destroying entire beaches, and even menacing cities \cite{Feagin, Martinez}. Therefore, maintaining the existing dune fields is a key issue for preserving national parks, beaches, and biodiversity, and a better understanding of the barchan-obstacle interactions is, thus, crucial for attaining this objective. On Mars, where the timescales are much greater than on Earth, the extrapolation of results obtained from experiments or simulations can give important hints about the ancient past of some of its barchan fields (or for predicting their far future).

Despite its importance for human activities and geophysics in general, the only investigations of dunes interacting with large obstacles are, to the best of our knowledge, those of \citeA{Bacik2} and \citeA{Assis4}. \citeA{Bacik2} investigated experimentally the behavior of two-dimensional (2D) dunes as they reached dune-size obstacles, conducting experiments in a narrow Couette-type circular channel. For different 2D obstacles obstructing the dune path (placed on the bottom wall), they found that there are two possible outcomes: dunes either cross over the obstacle or remain trapped. They proposed that the dune behavior is controlled by the fluid flow near the obstacle, which in its turn is disturbed by the size and shape of the obstacle. More recently, \citeA{Assis4} investigated the interaction of subaqueous barchans with dune-size obstacles by carrying out experiments in a water channel where a large 3D obstacle was fixed initially downstream of a single barchan. They varied the obstacle shape and size, the flow velocity, and the grains' properties, and found that subaqueous barchans can be blocked (trapped), bypass, or pass over the obstacles, with intermediate/transitional situations. Based on those results, \citeA{Assis4} proposed an \textit{ad hoc} classification map in which the interaction outcome depends on two dimensionless parameters: a modified Stokes number and the size ratio between the obstacle and the dune. In principle, those maps can be used for predicting the outcome of barchan-obstacle interactions.

Although previous studies showed the possible outcomes of the dune-obstacle interactions in subaqueous environment, and organized them in classification maps, the mechanisms involved in those interactions are still unknown. For example, the roles of flow disturbances, grains' inertia, and dune size remain to be investigated. One way of accessing the flow disturbances and forces involved are through numerical simulations. There are different models and techniques for simulating sand dunes, which can be classified basically in three types: sheared-continuum, continuum-continuum, and continuum-discrete methods. In the sheared-continuum methods, the granular bedform is considered a continuum medium that is put into motion by shearing stresses computed analytically \cite{Jackson_Hunt, Hunt_1, Weng}. Those methods were developed and exhaustively used for simulating eolian barchans \cite{Sauermann_4, Herrmann_Sauermann, Kroy_A, Kroy_C, Kroy_B, Schwammle, Parteli4}, with successful results because of the large size of dunes and number of grains involved. In the continuum-continuum models, both the fluid and grains are solved numerically using continuum equations. For example, \citeA{Sotiropoulos, Khosronejad, Khosronejad2} coupled large eddy simulation (LES) for the fluid (a water-suspension mixture) with a continuum model for the granular bed, and obtained good results for large subaqueous dunes. Finally, in continuum-discrete methods the fluid is solved using continuum equations, while the granular bedform is treated as a discrete medium. One example is the computational fluid dynamics - discrete element method (CFD-DEM), which solves the equations of motion for each solid particle (grain-scale resolution) within an Eulerian mesh for the fluid, being computationally more expensive than the other methods and usually limited to small bedforms. For instance, \citeA{Alvarez5, Alvarez7, Lima2} used this method (employing LES) to investigate 100-mm-long subaqueous barchans, obtaining detailed information of the disturbed fluid flow, trajectories of grains, and resultant forces for each grain (the latter not accessible from experiments).

In this paper, we inquire further into the interactions between barchans and dune-size obstacles in subaqueous environment. For that, we carried out numerical simulations where the fluid is solved at the grain scale by using LES and the motion of each grain is computed at all time steps by using discrete element method (DEM). With the outputs of simulations, we analyze the disturbances of the fluid flow, the trajectories of grains, and the resultant force on each grain, showing, for the first time, details of the dynamics during the interactions. Our results reproduce well the experiments reported in \citeA{Assis4}, showing that in some cases particles pass over (or touch) the obstacle, while in others they circumvent it (without touching it), or are even blocked (trapped), and those outcomes are in agreement with the classification map of that paper. We then show that the flow disturbances are in the origin of these outcomes, with a strong vortex between the lee face of the dune and the obstacle. This vortex results from the interactions of recirculation regions and horseshoe vortices, and has enough strength to deviate the main flow and carry grains around the obstacle in the bypass and trapped cases. In addition, we show the trajectories of individual grains and the distribution of the resultant force acting on each of them, which correlate well with the fluid trajectories of the disturbed flow. Our results represent a new step for understanding and predicting the different outcomes of dune-obstacle interactions.

\section{Materials and Methods}

Even if experiments can explain several aspects of the barchan-obstacle interaction, information such as forces acting on each grain and the dynamics of hidden (buried) grains are usually inaccessible. In order to overcome those drawbacks, we carried out CFD-DEM (computational fluid dynamics - discrete element method) simulations, following \citeA{Alvarez5, Alvarez7, Lima2, Lima3, Assis5}. CFD-DEM consists in an Euler-Lagrange approach, in which the fluid flow is computed in an Eulerian grid (CFD part) and the solid particles are tracked in a Lagrangian way (DEM part), and we used LES for CFD. LES computes the most energetic flow structures (those capable of entraining grains), while the smaller structures are estimated with sub-mesh models, optimizing the computational costs of simulations while maintaining accuracy of the fluid flow down to the grain scale \cite{Lima2}. In the subaqueous case, because of the relatively low inertia of solid particles with respect to water, grains follow closely the fluid flow, with meandering paths due to the different flow structures (including relatively small vortices, a few grain diameters in size). Therefore, LES is preferable than Reynolds-averaged Navier–Stokes (RANS) methods for simulating subaqueous dunes. However, if on the one hand LES recovers the flow structures down to the desired scales, on the other hand it becomes computationally expensive at large Reynolds numbers. In the eolian case, the use of RANS can be more adequate since the Reynolds numbers involved are much larger, and grains do not follow closely small vortices because of their higher inertia (compared to the subaqueous case).

For the computations, we made use of the open-source CFD code OpenFOAM (https://openfoam.org), open-source DEM code LIGGGHTS \cite{Kloss, Berger}, and open-source code \mbox{CFDEM} \cite<www.cfdem.com,>{Goniva} for coupling the CFD and DEM parts. Next, we describe briefly the equations used and the numerical setup, more details are available in \citeA{Lima2} and in the Supporting Information.

\subsection{Model}

We solve the problem using an Euler-Lagrange framework, with 4-way coupling and second-order (two-fluid) models for the fluid \cite{Giudice}. The Lagrangian part (computed here using DEM) solves the equations of linear and angular momentum (Equations \ref{Fp} and \ref{Tp}, respectively) for each solid particle (grain),

\begin{equation}
	m_{p}\frac{d\vec{u}_{p}}{dt}= \vec{F}_{fp} + \vec{F}_{c} + m_{p}\vec{g}\, ,
	\label{Fp}
\end{equation}

\begin{equation}
	I_{p}\frac{d\vec{\omega}_{p}}{dt}=\vec{T}_{c}\, ,
	\label{Tp}
\end{equation}

\noindent where $\vec{g}$ is the acceleration of gravity and, for each grain, $m_{p}$ is the mass, $\vec{u}_{p}$ is the velocity, $I_{p}$ is the moment of inertia, $\vec{\omega}_{p}$ is the angular velocity, $\vec{F}_{c}$ is the resultant of contact forces between solids, $\vec{F}_{fp}$ is the resultant of fluid forces acting on the solids, and $\vec{T}_{c}$ is the resultant of contact torques between solids. We note that the right-hand side of Equation \ref{Fp} corresponds to the resultant force $\vec{F}_p$ on each particle, and that in Equation \ref{Tp} we neglect torques caused directly by the fluid, since those due to contacts are much higher \cite{Tsuji, Tsuji2, Liu}.

For the contact forces, $\vec{F}_{c}$, and torques, $\vec{T}_{c}$, we consider the particle-particle and particle-wall contacts,

\begin{equation}
	\vec{F}_{c} = \sum_{i\neq j}^{N_c} \left(\vec{F}_{c,ij} \right) + \sum_{i}^{N_w} \left( \vec{F}_{c,iw} \right)
	\label{Fc}
\end{equation}

\begin{equation}
	\vec{T}_{c} = \sum_{i\neq j}^{N_c} \vec{T}_{c,ij} + \sum_{i}^{N_w} \vec{T}_{c,iw}
	\label{Tc}
\end{equation}

\noindent where $\vec{F}_{c,ij}$ and $\vec{F}_{c,iw}$ are the contact forces between particles $i$ and $j$ and between particle $i$ and the wall, respectively, $\vec{T}_{c,ij}$ is the torque due to the tangential component of the contact force between particles $i$ and $j$, and $\vec{T}_{c,iw}$ is the torque due to the tangential component of the contact force between particle $i$ and the wall. $N_c$ - 1 is the number of particles in contact with particle $i$, and $N_w$ is the number of particles in contact with the wall. We consider a Hertzian model in Equations \ref{Fc} and \ref{Tc}, in which contact forces are decomposed into normal and tangential components, as shown briefly in the Supporting Information and in more details in \citeA{Lima2}.

For the resultant fluid force acting on each grain, $\vec{F}_{fp}$, we consider the fluid drag, $\vec{F}_{d}$, the force due to pressure gradient, $\vec{F}_{press}$, the force due to the deviatoric stress tensor, $\vec{F}_{\tau}$, and the virtual mass force, $\vec{F}_{vm}$, as shown in Equation \ref{Ffp_sim},

\begin{equation}
	\vec{F}_{fp} = \vec{F}_{d} + \vec{F}_{press} + \vec{F}_{\tau} + \vec{F}_{vm} \, ,
	\label{Ffp_sim}
\end{equation}

\noindent and we do not take into account the Basset, Saffman, and Magnus forces since they are considered negligible in CFD-DEM simulations \cite{Zhou}. The details for computing each of the considered forces are described in the Supporting Information.

The Eulerian part (computed here using CFD) solves the equations of mass (Equation \ref{mass}) and momentum (Equation \ref{mom}) for the fluid, in which we use an unresolved approach. For that, the equations are phase-averaged in a volume basis while assuring mass conservation:

\begin{equation}
	\frac{\partial \left( \alpha_{f} \rho_{f} \right)}{\partial t} + \nabla \cdot \left ( \alpha_{f} \rho_{f} \vec{u}_{f} \right ) = 0 \,\,,
	\label{mass}
\end{equation}

\begin{equation}
	\frac{\partial \left ( \alpha_{f} \rho_{f} \vec{u}_{f} \right ) }{\partial t} + \nabla \cdot \left ( \alpha_{f} \rho_{f} \vec{u}_{f} \vec{u}_{f} \right ) = -\alpha_{f} \nabla P - \vec{f}_{exch} + \alpha_{f} \nabla \cdot  \vec{\vec{\tau}}_{f}  + \alpha_{f} \rho_{f} \vec{g} \,\,,
	\label{mom}
\end{equation}

\noindent where $\vec{u}_{f}$ is the fluid velocity, $\rho_{f}$ is the fluid density, $\alpha_{f}$ is the volume fraction of the fluid, $P$ the fluid pressure, $\vec{\vec{\tau}}$ the deviatoric stress tensor of the fluid, and $\vec{f}_{exch}$ is an exchange term between the grains and the fluid,

\begin{equation}
	\vec{f}_{exch} = \frac{1}{\Delta V}\sum_{i}^{n_{p}} \left( \vec{F}_{d} +  \vec{F}_{vm} \right) \,\, .
	\label{forces_exchange}
\end{equation}

\noindent where $n_p$ is the number of particles in a considered cell of volume $\Delta V$. Because the forces due to the pressure gradient and deviatoric stress tensor were split from the remaining fluid-particle forces during the averaging process (for obtaining Equations \ref{mass} and \ref{mom}), $\vec{f}_{exch}$ is different from $n_p \vec{F}_{fp} / \Delta V$. As used in our LES simulations, Equations \ref{mass} and \ref{mom} correspond to filtered quantities, and the deviatoric stress term is split into viscous and turbulent terms, as shown in the Supporting Information. In addition, details on the computation of the drag and virtual mass forces are also available in the Supporting Information.

\subsection{Numerical setup}

\begin{figure}[ht]
	\begin{center}
		\includegraphics[width=0.9\linewidth]{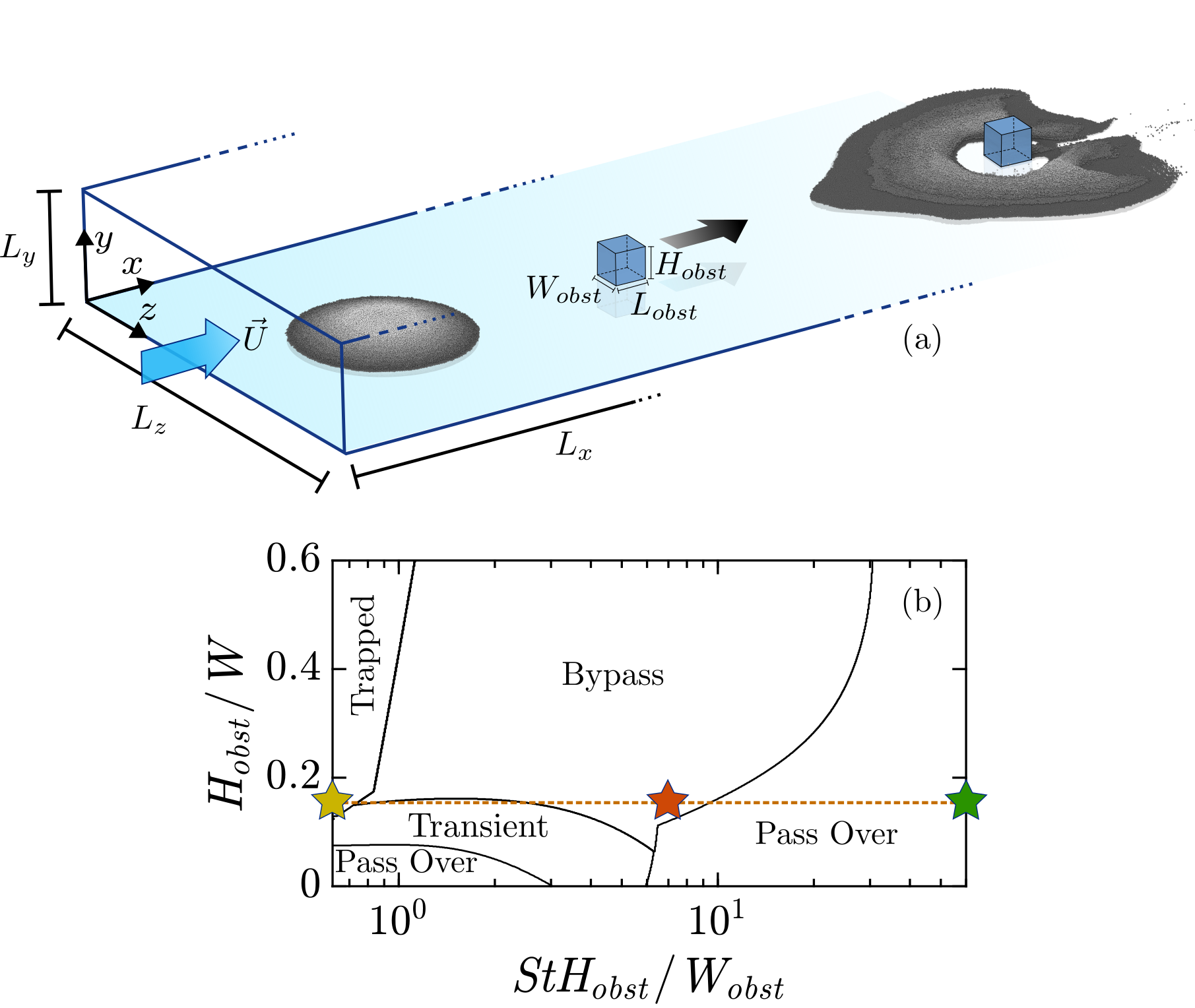}\\
	\end{center}
	\caption{(a) Layout of the numerical setup, showing the channel dimensions, the flow direction, the initial pile, the obstacle, and the dune and obstacle at a posterior time (the obstacle remained in the same place, in the figure it is displaced for visualization purposes only). In all cases, the upstream pile was initially placed at 3 cm from the CFD inlet. (b) Simulated cases in the interaction map proposed by \citeA{Assis4}: size ratio vs. Stokes effect, i.e., $H_{obst}/W$ vs. $St H_{obst} / W_{obst}$ \cite<Figure modified from>{Assis4}.}
	\label{fig:setup}
\end{figure}

The CFD domain is a channel $L_x$ = 0.4 m long, $L_y$ = $\delta$ = 0.025 m high, and $L_z$ = 0.1 m wide, with $x$, $y$ and $z$ being the longitudinal \corr{(streamwise)}, vertical and spanwise directions, respectively (in dimensionless terms, $L_x/H_{obst}$ $=$ 80, $L_y/H_{obst}$ $=$ 5, and $L_z/H_{obst}$ $=$ 20, where $H_{obst}$ is the obstacle height). The boundary conditions for the CFD domain are periodic conditions in the longitudinal and spanwise directions, no-slip conditions on the bottom wall \corr{(which was a smooth wall)}, and free slip on the top boundary ($y$ = $\delta$). The latter condition is imposed because $L_y$ = $\delta$ corresponds to the channel half height in the experiments reported in \citeA{Assis4}, the real channel height being 2$\delta$ = 0.050 m; with that, computational costs are reduced. The obstacle was built as a boundary condition of the bottom wall, centered on the middle of the bottom wall (in both the streamwise and spanwise directions) to minimize entrance effects. The obstacles were rectangular prisms, with dimensions $L_{obst}$ $=$ 5 mm, $H_{obst}$ $=$ 5 mm and $W_{obst}$ $=$ 5 mm for the bypass, $L_{obst}$ $=$ 0.5 mm, $H_{obst}$ $=$ 5 mm and $W_{obst}$ $=$ 0.5 mm for the pass over, and $L_{obst}$ $=$ 5 mm, $H_{obst}$ $=$ 5 mm and $W_{obst}$ $=$ 70 mm for the trapped case, where \corr{$L_{obst}$ is the streamwise length and $W_{obst}$ is the width (spanwise length) of the obstacles (shown in Figure \ref{fig:setup})}. \corr{The use of periodic conditions in the spanwise direction corresponds to an infinite array of obstacles placed side by side. This condition can affect results in the trapped case, given the width of the obstacle. In this case, the free spanwise distance between obstacles is 30 mm, which can create channeling and accelerate particles more than in an unbounded case. However, in the experiments of \citeA{Assis4} the distance between the obstacle and each vertical wall was 40 mm, close to the free distance in simulations. In addition, the width of the wider obstacle (trapped case) corresponds to 70\% of the channel width, so that we do not expect significant effects from lateral walls \cite{Franklin_8, Alvarez3}, or, similarly, from using periodic conditions in the spanwise direction}.

In the simulations, the cross-sectional mean velocity is $U$ $\approx$ 0.278 m/s, the channel Reynolds number based on $U$ and the channel height, Re = $U 2\delta \nu^{-1}$, is 13,900, the Reynolds number based on shear velocity $u_*$ and the channel half height, Re$_*$ = $u_* \delta \nu^{-1}$, is 400, the Reynolds number based on $U$ and the obstacle height, Re = $U H_{obst} \nu^{-1}$, is 1,390, the Stokes number based on the mean grain diameter $d$ \cite{Andreotti_6}, $St \,=\, U d \rho_p / (18\mu)$, is 8, and the Shields number, $\theta$ = $(\rho_f u_*^2)/((\rho_p - \rho_f )gd)$, is 0.09, where $\mu$ and $\nu$ are the dynamic and kinematic viscosities of the water, respectively, and $\rho_p$ the density of the solid material (glass). The solid particles of the initial heap consisted of 10$^5$ glass spheres, with sizes following a Gaussian distribution (randomly selected once) within 0.15 mm $\leq$ $d_p$ $\leq$ 0.25 mm (mean value equal to 0.2 mm and standard deviation equal to 0.025 mm). The exact distribution of grain diameters as well as the mechanical properties of the used particles are listed in Table \ref{diameterAndProps}. The boundary conditions for the grains were solid wall at the bottom boundary, free exit at the outlet, and no grain influx at the inlet, so that the number of grains in the domain decreased along time \cite<as in our previous experiments,>{Assis, Assis2, Assis4}.

For dealing with turbulence close to the wall, we made use of the wall-adapting local eddy-viscosity (WALE) model \cite{Nicoud} along with a cube-root volume-based approach to determine the width of the LES filter. We did not use wall functions, as our mesh resolution ensured that $y^+$ $=$ $yu_*/\nu$ remained close to 1 near the wall, allowing us to resolve the near-wall turbulence without requiring additional modeling assumptions. Besides, the WALE model itself incorporates a damping mechanism that accounts for regions of high strain and low $y^+$, making an explicit damping function unnecessary. This approach allowed us to capture all re-circulations and high-energy vortices necessary in the subaqueous bedload transport. Regarding the spatial discretization, all mesh elements were hexahedral and orthogonal, and, in each simulation, it was necessary to remove the control volumes corresponding to the obstacle, leading to a distinct number of control volumes in each case. Additionally, a localized refinement was applied around the obstacles to accurately capture vortex structures and maintain a given target $y^+$ value in their vicinity (we chose 1.02, so that smaller mesh elements could encompass at least one grain). The numbers of control volumes $N_{cv}$, numbers of divisions of the domain in the $x$, $y$ and $z$ directions, vertical position of the center of the first control volume  $y^{+}_{1st}$, and average height (size) of the used meshes close to the bottom wall $y^{+}_{avg}$, for each simulated case, are listed in Table \ref{meshProps}. Images of the meshes and more information on the numerical schemes used are available in the Supporting Information. Because in our computations the size of grains is comparable to that of CFD meshes, we used a diffusion equation for the void fraction $\alpha_f$ and coupling forces $\vec{f}_{exch}$,

\begin{equation}
	\frac{\partial \xi}{\partial t} = \frac{\lambda}{\Delta t_{CFD}}\nabla^2\xi \,\,,
	\label{diffusion_appendix}
\end{equation}

\noindent where $\lambda$ is the characteristic smoothing length (we imposed $\lambda$ = 3$d$ in the present simulations), $\Delta t_{CFD}$ is the time step of CFD computations, and $\xi$ = $\alpha_f$ or $\vec{f}_{exch}$. The current LES-DEM setup was extensively tested and compared with experiments in \citeA{Lima2} for the case of isolated barchans, so that a complete description of parameters, meshes, convergence, tests, and validation can be found in \citeA{Lima2}. More details of the numerical setup are also available in the Supporting Information.

\begin{table}[!h]
	\centering
	\caption{For each of the simulated cases, numbers of control volumes $N_{cv}$, numbers of divisions of the domain in the $x$, $y$ and $z$ directions ($d_x$, $d_y$ and $d_z$, respectively), vertical position of the center of the first control volume  $y^{+}_{1st}$ scaled in inner-wall units ($\nu/u_*$), and average height (size) of the used meshes close to the bottom wall $y^{+}_{avg}$.}
	\begin{tabular}{ccccccc}
			\hline
			Case & $N_{cv}$ & $d_x$ & $d_y$ & $d_z$ & $y^{+}_{1st}$ & $y^{+}_{avg}$\\ \hline
			No obstacle & 450000 & 250 & 30 & 60  & 1.02 & 10.6  \\
			Bypass & 448000 & 250 & 30 & 60 & 1.02 & 10.6  \\
			Pass over & 449920 & 250 & 30 & 60 & 1.02 & 10.6  \\
			Trapped & 441600 & 250 & 30 & 60 & 1.02 & 10.6 \\
			\hline
	\end{tabular}
	\label{meshProps}
\end{table}

\begin{table}[!h]
	\centering
	\caption{Distribution of diameters for the grains in the initial pile, where $N_{d}$ is the number of grains for each diameter $d$, and physical properties.}
	\begin{tabular}{cc|lc}
		\hline
		$N_{d}$ & Diameter $d$ ($mm$) & \multicolumn{2}{l}{Properties} \\ \hline
		2276       & 0.15  &   Sliding Friction Coeff. $\mu$   & 0.6 \\
		13592      & 0.175  &  Rolling Friction Coeff. $\mu_r$  & 0.0 \\
		68267      & 0.2    &  Restitution Coef. $e$   & 0.1\\
		13591      & 0.225  &  Poisson Ratio $\sigma$   & 0.45   \\
		2274       & 0.25   &  Young's Modulus $E$ (MPa) & 5    \\
		$\cdots$       & $\cdots$   &  Density $\rho_p$ (kg/m$^3$)  & 2500  \\
		\hline		
	\end{tabular}
	\label{diameterAndProps}
\end{table}

Prior to simulations of the barchan-obstacle interactions, we carried out an LES simulation of a pure water flow in the periodic channel (with the obstacle), until reaching a fully-developed turbulent flow. \corr{A study of mesh convergence of averaged profiles was carried out by \citeA{Lima2} in a similar channel (but without the obstacle), which shows that the mean velocity profile of the base flow corresponds to a fully-developed channel flow in the hydraulic-smooth regime (typical law of the wall).} The results were stored to be used as initial condition for the fluid in the LES-DEM simulations. The next step was to let the grains settle by free fall in still water, forming one conical pile with radius $R$ $\approx$ 0.0145 m, height $h$ $\approx$ 0.003 m, and distant 2$R$ in the longitudinal direction from the obstacle. Finally, the last step was to impose the turbulent flow (stored in a previous step) in the channel, in the presence of the initial pile and obstacle, as shown in Figure \ref{fig:setup}a. \corr{In our unresolved CFD-DEM computations, the presence of a bed of particles is modeled by locally reducing the porosity in the cells containing particles, introducing an additional resistance to the fluid flow. This artifice allows the code to run even if particles are relatively large with respect to the cells. The fluid velocity is thus computed throughout the entire domain, including the region occupied by the particles, and naturally decreases in zones with higher particle concentration due to the reduced porosity. Therefore, there is a mismatch between the initial field (pure water flow) and the new hydrodynamic equilibrium imposed by the presence of the granular pile, but the solver quickly corrects this within the first time steps. The used procedure} assured that turbulence generation was kept along simulation, while the relative large extension of the channel with respect to the dune and obstacle sizes avoided the recirculation regions and reattachment points from reaching the channel inlet. In all simulated cases, the time step for the DEM was 2.5 $\times$ 10$^{-6}$ s, which corresponds to less than 20\% of the Rayleigh time \cite{Derakhshani}, and that of CFD was 2.5 $\times$ 10$^{-4}$ s, which respects the CFL (Courant-Friedrichs-Lewy) criterion \cite{Courant}. \corr{Although the implicit scheme used is unconditionally stable from a numerical standpoint, CFL condition remains highly relevant in practice -- not for stability reasons, but for ensuring sufficient temporal resolution and accuracy in transient-flow simulations (see Supporting Information for a brief explanation of the numerical schemes used).} Convergence tests were carried out and published in \citeA{Lima2}. A movie showing this complete simulation is available in the Supporting Information.

We simulated three cases observed by \citeA{Assis4}: the bypass, pass over, and trapped. They are shown in Figure \ref{fig:setup}b in the $H_{obst}/W$ (size ratio) vs. $St H_{obst} / W_{obst}$ (modified Stokes number) diagram proposed by \citeA{Assis4}, where $W$ is the barchan width.

\section{Results and discussion}

\subsection{Morphology}

\begin{figure}[ht]
	\begin{center}
		\includegraphics[width=0.7\linewidth]{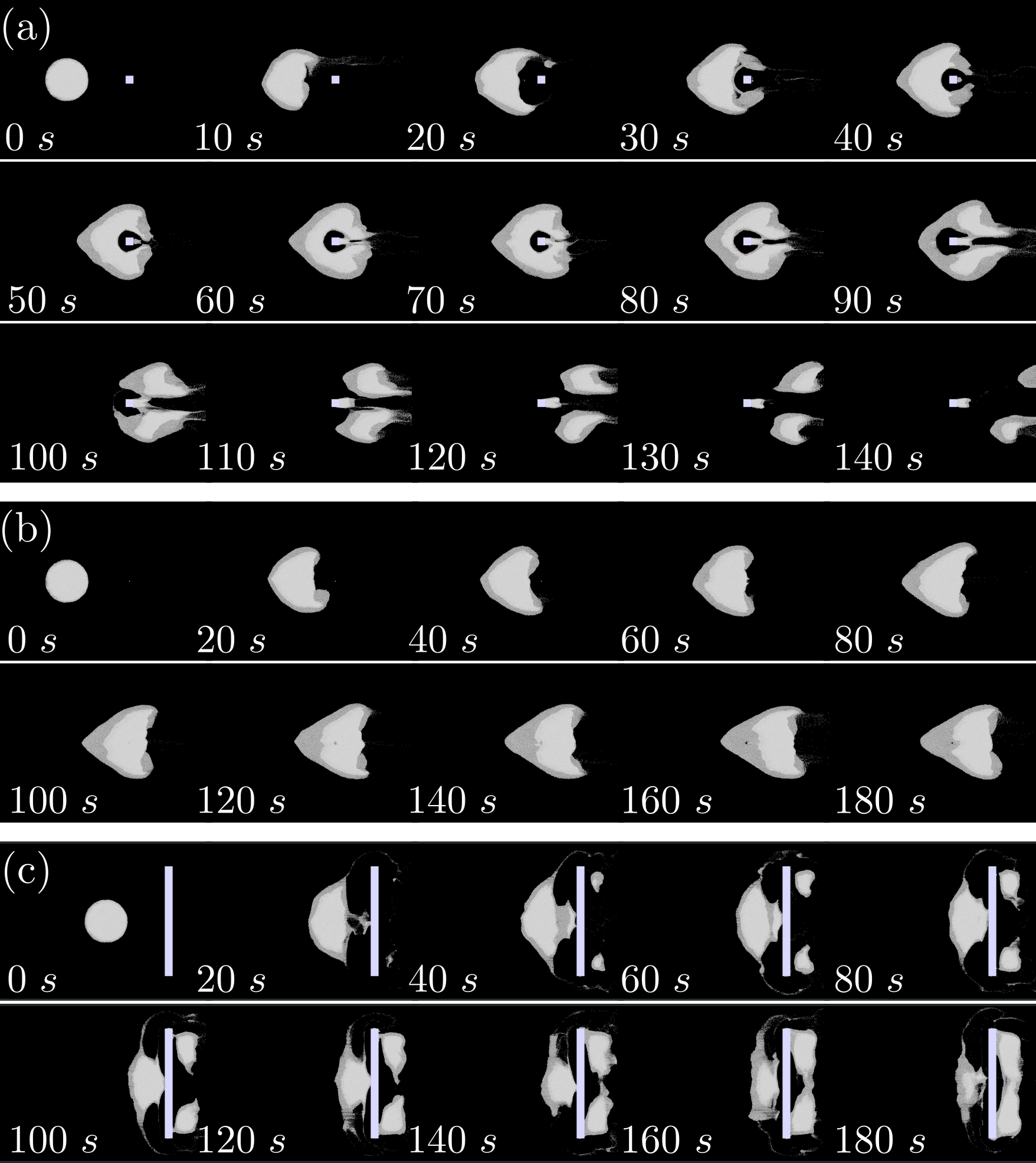}\\
	\end{center}
	\caption{Snapshots showing top view images of a dune interacting with the obstacle for the (a) bypass, (b) pass over, and (c) trapped cases. The time instants are shown on the bottom left of each snapshot.}
	\label{fig:morphology}
\end{figure}

Once the water flow is imposed in the computational domain, the conical pile begins to deform into a barchan dune, including the formation of a well defined avalanche face and recirculation region, until reaching a region close to the obstacle. In the case of the largest obstacle, the initial pile has not enough time (or distance) to completely develop into a crescent shape before interacting with the it (as can be seen in Figure \ref{fig:morphology}c, discussed next).

Figures \ref{fig:morphology}a, \ref{fig:morphology}b and \ref{fig:morphology}c show top-view images of a dune interacting with the obstacle for the bypass, pass over, and trapped cases, respectively. We first observe that the agreement with images from the experiments presented in \citeA{Assis4} is good, with the dune circumventing the obstacle in the bypass case, touching the obstacle and going over it in the pass over case, and stretching in the transverse direction (and being destroyed) in the trapped case. For the latter, most of grains are trapped in the recirculation region (93\% of the total), while in the experiments of \citeA{Assis4} the number of grains entrained further downstream seems a little higher. The good agreement indicates that the methodology and numerical setup used in our simulations capture well the dynamics at the bedform scale. Therefore, we analyze in the following \corr{the fluid flow, and} the trajectories and forces at the grain scale, something for which the simulations are more accessible than experiments (in the case of forces, simulations are currently the only option). Movies showing the motion of grains during the interactions of Figures \ref{fig:morphology}a, \ref{fig:morphology}b and \ref{fig:morphology}c are available in the Supporting Information.

\subsection{\corr{Fluid flow}}

\begin{figure}[ht]
	\begin{center}
		\includegraphics[width=0.99\linewidth]{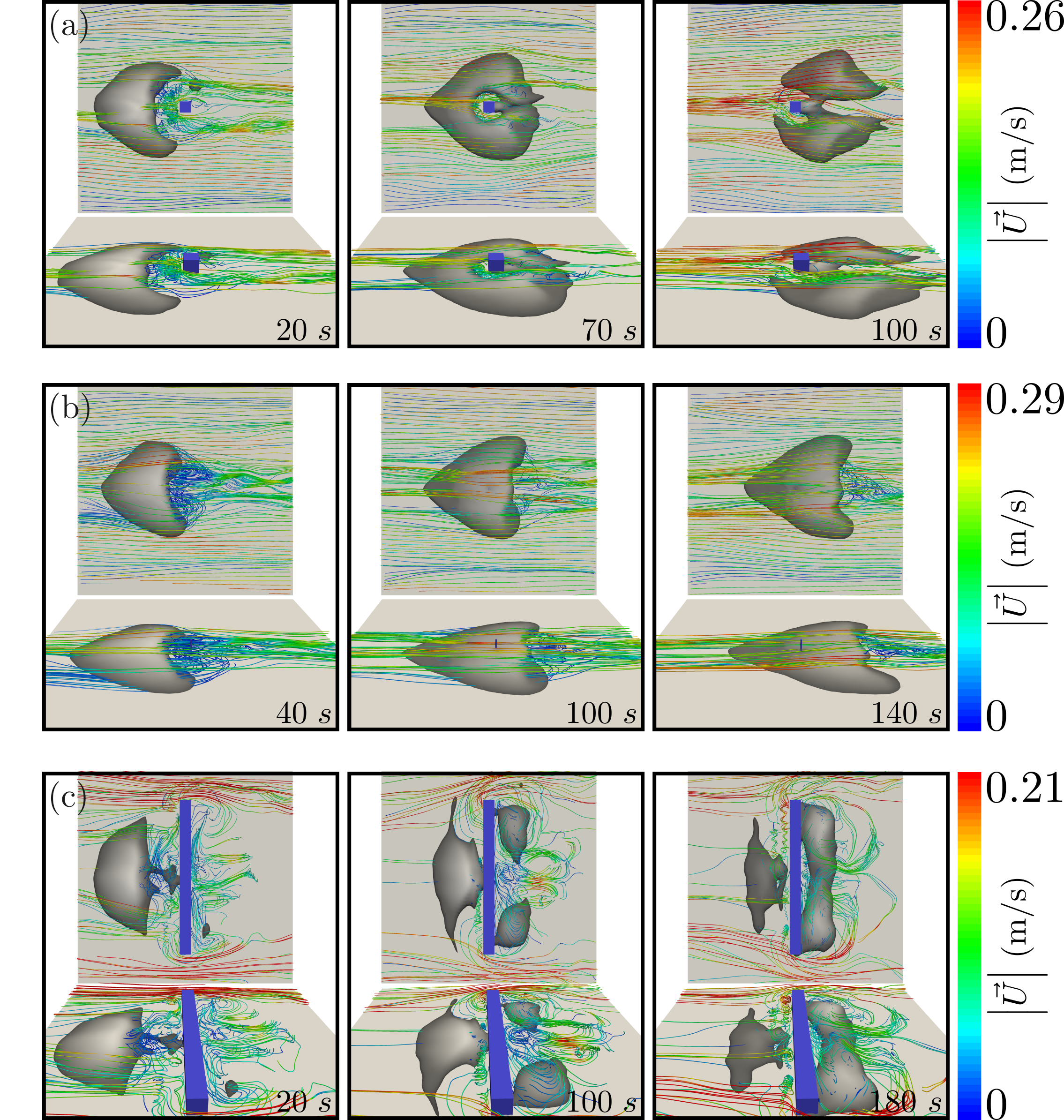}\\
	\end{center}
	\caption{Streamlines (instantaneous) at different stages of the barchan-obstacle interaction, for the (a) bypass, (b) pass over, and (c) trapped cases. In each panel, the upper row shows top view and the lower row perspective images, and the colorbar on the right corresponds to the magnitude of the velocity.}
	\label{fig:streamlines}
\end{figure}

\corr{We inquire next into} the fluid flow, which is the mechanism of grain entrainment, and which we computed with a spatial resolution of the order of the grain diameter in the region close to the dune surface. There is a vast literature for the channel or boundary-layer flow disturbed by a prismatic obstacle, in particular cubes, in the absence of grains \cite{Martinuzzi, Hussein, Hwang, Lim, Launay, Diaz}. Those studies investigated, experimentally and numerically, the critical points within the flow, the spectrum of turbulent energy, and instabilities, identifying the recirculation regions and vortices, vortex shedding, and disturbances of the flow around recirculation regions. \corr{\citeA{Baines} was perhaps the first to show that the formation of a recirculation vortex (a horseshoe vortex) in front of prismatic obstacles depends on their aspect ratio and also on the incoming velocity profile. They showed that typical boundary layers (law of the wall) deviating from cubes and very wide prisms (corresponding to the bypass and trapped cases in our tests) produce a horseshoe vortex whose size and intensity increase with the velocity gradient of the incoming flow. For uniform flows, they showed that the horseshoe vortex is virtually absent. Later, \citeA{Castro} confirmed that the incoming flow is important, with the size of the horseshoe vortex increasing with the boundary-layer thickness, and \citeA{Martinuzzi} showed that for high obstacles a main vortex is formed close to their upstream face, with a series of small vortices upstream the main one. In general, those works show that close to the obstacle a turbulent boundary layer} is highly disturbed, with, in particular, the presence of an upstream horseshoe vortex that deflects in the transverse direction a great portion of the fluid flow \corr{(its size depending both on the incoming flow and obstacle geometry)}. When the cubical obstacle is over a granular bed, the disturbed flow erodes the bed with more intensity close to the upstream corners, as shown by \citeA{Tominaga} in wind-tunnel experiments and CFD simulations, in a direct effect of flow disturbances.

Because of the large quantity of data, we present next typical streamlines (instantaneous) at different stages of the barchan-obstacle interaction. Figures \ref{fig:streamlines}a-c show streamlines at three instants for the bypass, pass over, and trapped cases, respectively, and in each panel the upper row shows top view images and the lower row images in perspective. The colorbar on the right corresponds to the magnitude of the velocity.

In the bypass case (Figure \ref{fig:streamlines}a), we observe at $t$ = 20 s a strong vortex between the dune and the obstacle. Because a recirculation region exists just downstream the dune crest \cite{Bagnold_1, Kroy_A, Kroy_B, Lima2} and a horseshoe vortex just upstream the obstacle \cite{Martinuzzi, Hwang, Launay, Diaz}, the observed vortex results from the interaction between the recirculation and the horseshoe vortices (please see Figure S3 of Supporting Information for pathlines of the flow around the obstacle only, i.e., without grains). The resulting vortex is strong enough for pushing the grains away from the obstacle, making them circumvent the obstacle instead of passing over it. The same vortex is still present at $t$ = 70 s, and, at $t$ = 100 s, when there is no longer a bedform upstream the obstacle, the horseshoe vortex hinders grains on the flanks of the obstacle from touching it. A similar behavior is observed for the trapped case (Figure \ref{fig:streamlines}c), with the exception that at $t$ = 180 s a bedform is still upstream the obstacle (so that the interaction between vortices is still present). We note also that there is a suction region in the center of the upstream face. This suction is due to longitudinal streamlines that reach the center of the upstream wall, since, given the large width of the obstacle, the flow must split in two parts in that region. For the pass over case, the width of the prism is so small that its horseshoe vortex is not strong enough for, from the interaction with the barchan recirculation region, generating a vortex that pushes away the grains.

\begin{figure}[ht]
	\begin{center}
		\includegraphics[width=0.80\linewidth]{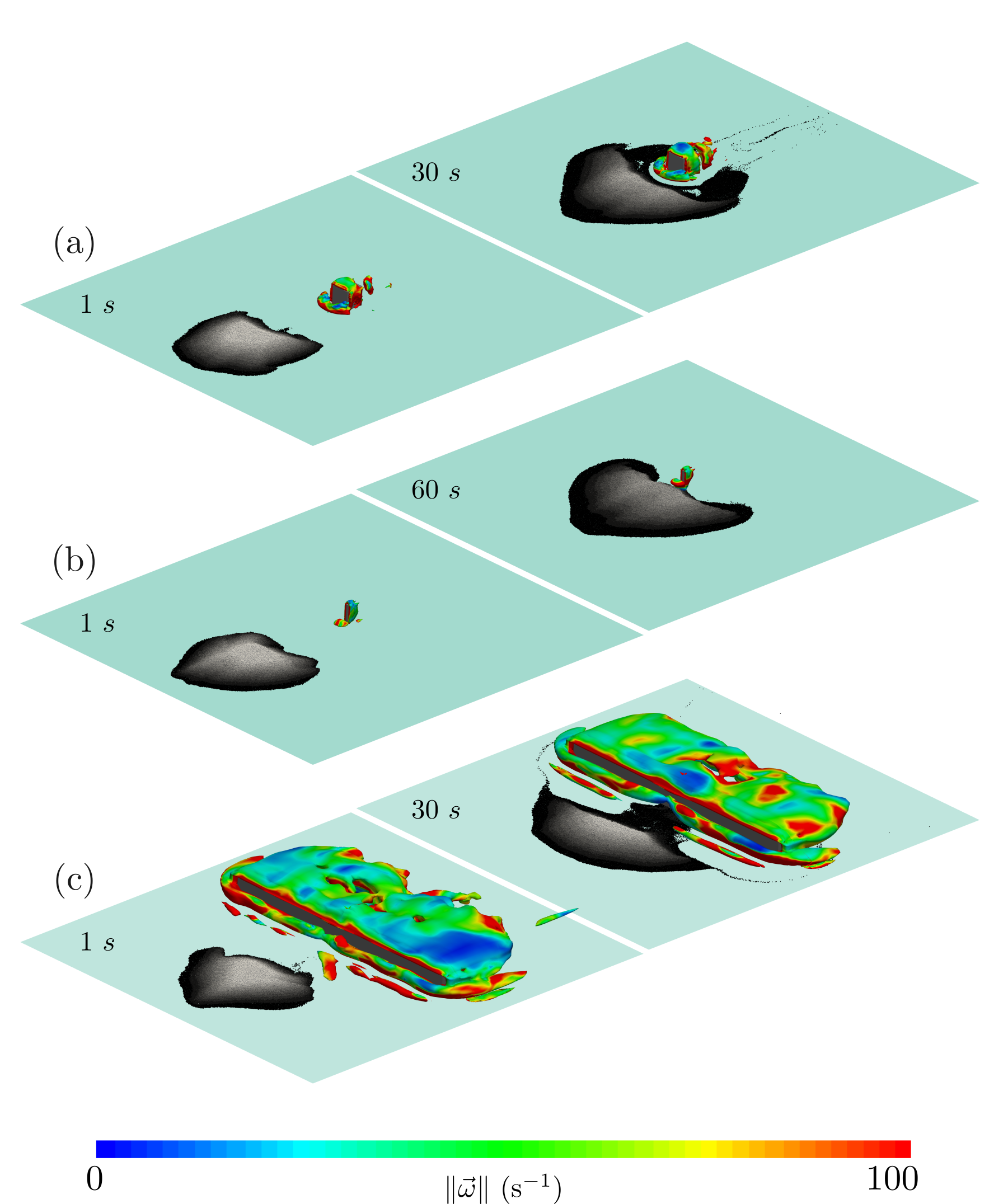}\\
	\end{center}
	\caption{Visualization of vortices and grains in the (a) bypass, (b) pass over, and (c) trapped cases. The vortices are visualized using the $Q$-criterion with $Q$ $=$ 100, times are indicated on the panels, and the colorbar corresponds to the magnitude of the vorticity.}
	\label{fig_Q_vortices}
\end{figure}

\begin{figure}[ht]
	\begin{center}
		\includegraphics[width=0.70\linewidth]{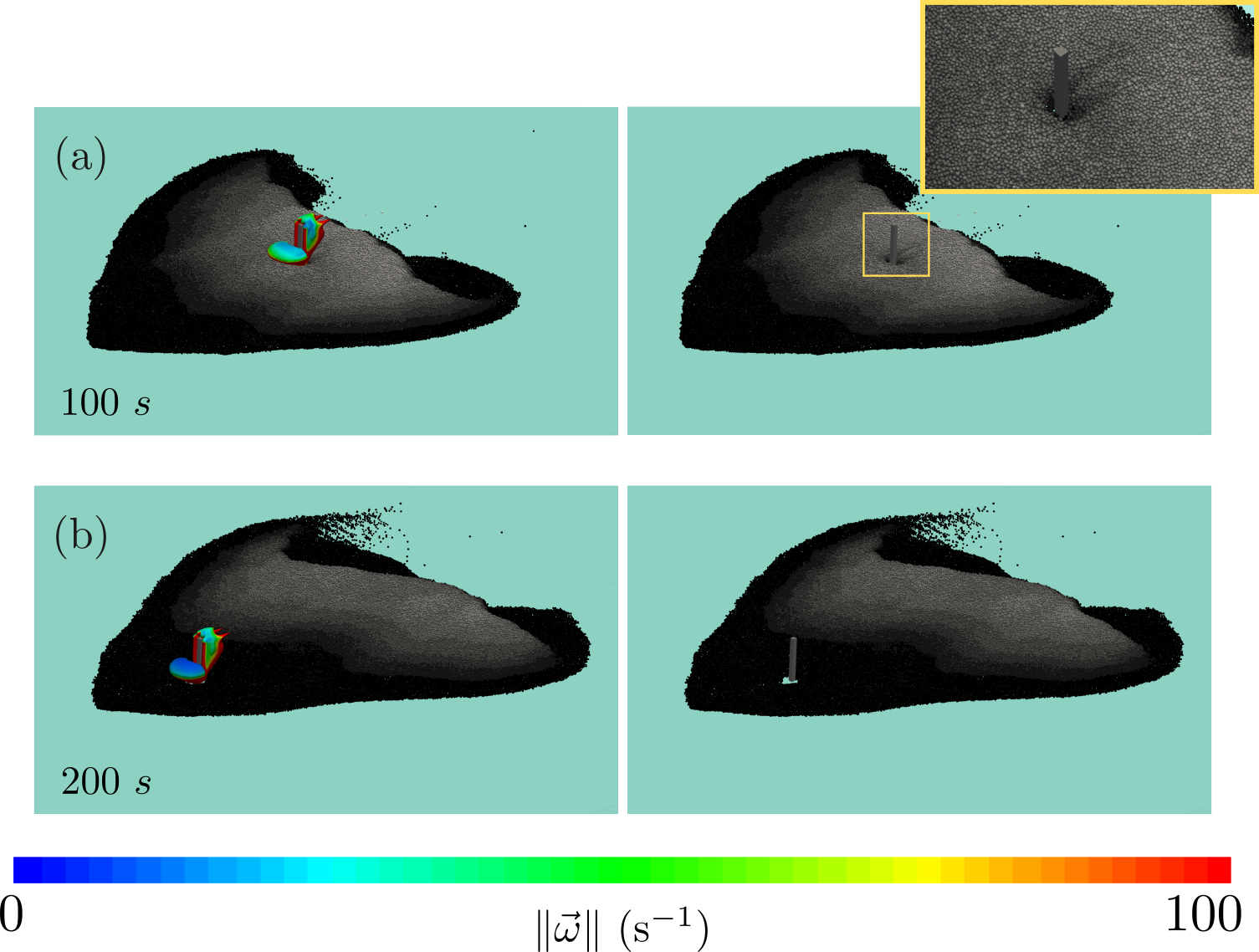}\\
	\end{center}
	\caption{Visualization of vortices and grains (left) and only grains (right) for the pass over case, in two different instants: (a) $t$ $=$ 100 s; (b) $t$ $=$ 200 s. The vortices are visualized using the $Q$-criterion with $Q$ $=$ 100, times are indicated on the panels, and the colorbar corresponds to the magnitude of the vorticity.}
	\label{fig_Q_vortices2}
\end{figure}

In order to visualize and further inquire into the vortical structures between the bedform and the obstacle, we adopted the $Q$-criterion \cite{Dubief},

\begin{equation}
	Q = \frac{1}{2} \left(|\vec{\vec{R}}|^{2} - |\vec{\vec{S}}|^{2} \right) \,\,,
	\label{eq_Q}
\end{equation}

\noindent where $\vec{\vec{R}}$ is the rotation and $\vec{\vec{S}}$ the deformation tensor, so that $Q$ is the second invariant of the velocity gradient tensor $\nabla \vec{u}_f$. Figures \ref{fig_Q_vortices2}a-c show visualizations of the grains and vortices using the $Q$-criterion for the bypass, pass over, and trapped cases, respectively. We used $Q$ $=$ 100 and the colors in the colorbar correspond to the magnitude of the vorticity. We observe that, indeed, the resulting vortices are much stronger in the bypass and trapped cases with respect to the pass over case, molding the form of the dune as it interacts with the obstacle. Because the vortices in Figure \ref{fig_Q_vortices2}b are relatively small and the dune touches the obstacle in future times, we plotted the pass over case at $t$ $=$ 100 s and $t$ $=$ 200 s in Figures \ref{fig_Q_vortices2}a and \ref{fig_Q_vortices2}b, respectively. On the left of each figure, we superposed the vortices and the grains, and on the right we plotted only the grains. We observe that at $t$ $=$ 100 s, when the core of the barchan is passing the obstacle (Figure \ref{fig_Q_vortices2}a), some of its grains are in contact with the thin prism (which is higher than the dune), submerging its lower part. At $t$ $=$ 200 s, when the barchan has almost passed the obstacle, a monolayer of grains remains around the barchan's toe \cite<also observed in experiments,>{Alvarez4, Assis2, Assis3}. Because the number of grains in the monolayer is small and most of the thin prism becomes exposed to the fluid flow, the horseshoe vortex forms again and is strong enough to deviate the monolayer of grains from the region near the prism, creating a small void region (right panel of Figure \ref{fig_Q_vortices2}b).

We note that in the subaqueous case the density ratio between the grains and the fluid is relatively small ($S$ $=$ 2.5, while $S$ $\approx$ 2500 in the eolian case), so that grains moving as bedload roll and slide, following closely the fluid flow. Therefore, the existence of a strong vortex around the obstacle and a fluid flow that accelerates around it pushes the grains away: the grains follow paths that are closely related with the fluid flow, avoiding the obstacle. The situation is different in the eolian case, where grains have considerable inertia and move by saltation (ballistic flights), tending to follow straight lines. In that case, grains can easily reach the obstacle even if the fluid flow deviates from it. For instance, we frequently observe the accumulation of sand upstream houses and other human built structures in desert areas. However, for relatively high geometries such as the pyramids of Mero\"e in Sudan, scour just upstream the obstacle is observed, associated with the presence of a horseshoe vortex \cite{Raffaele}. In the specific case of barchan-obstacle interactions, the void regions found in the subaqueous bypass and trapping patterns are not expected to be so strong in the eolian as in the aquatic environment. Therefore, even if our results represent a new step further for understanding dune-obstacle interactions, extrapolations to eolian environment must be carried out with caution.

\subsection{Trajectories of grains and resultant forces}

\begin{figure}[ht]
	\begin{center}
		\includegraphics[width=0.7\linewidth]{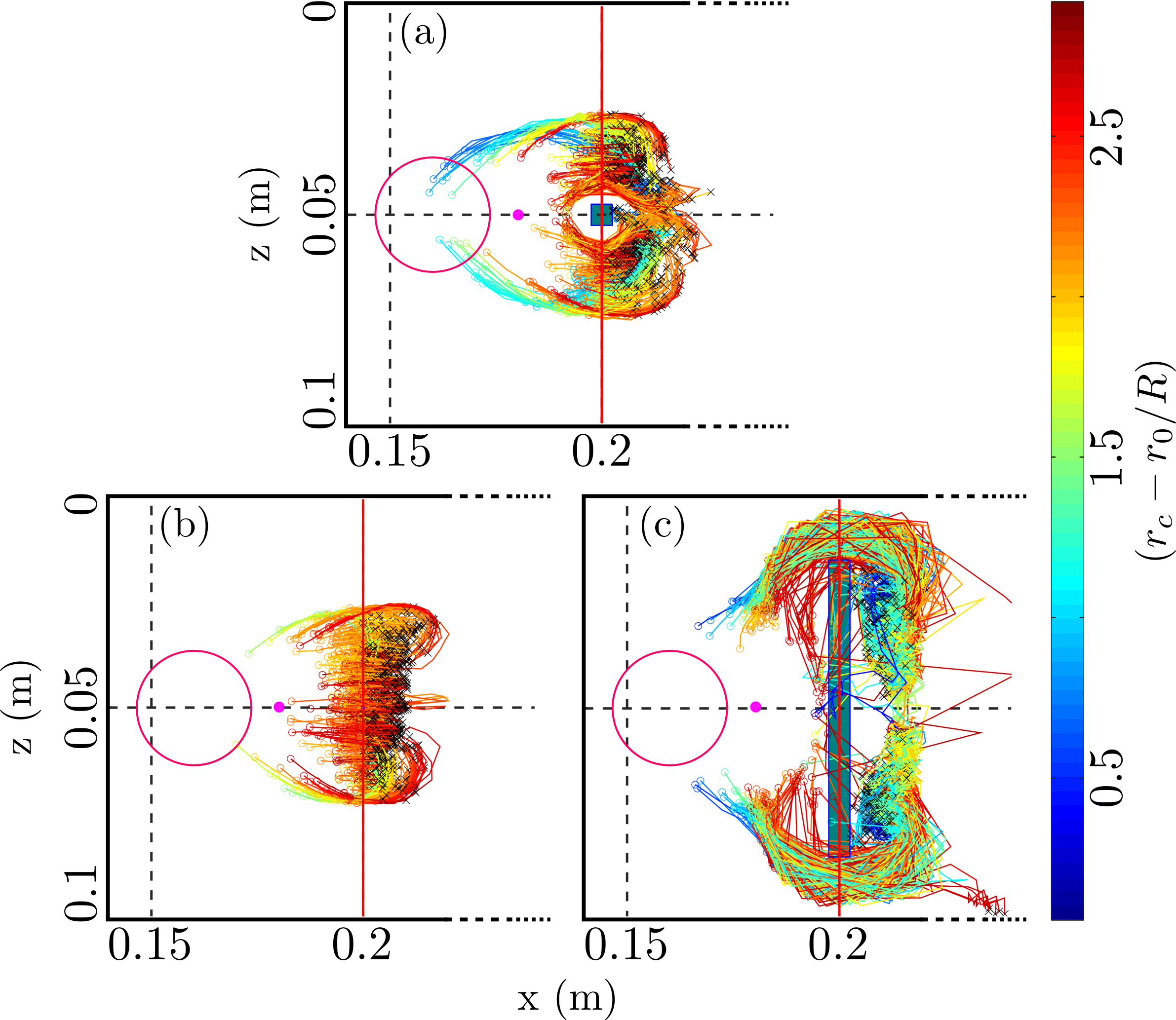}\\
	\end{center}
	\caption{Trajectories of individual grains for the (a) bypass, (b) pass over and (c) trapped cases. In the panels, the small circles represent the starting point of the considered particle, the black crosses represent its final position, the large circle is the mean perimeter of the initial pile, the vertical red line represents the cross section passing by the center of the obstacle, and the pink-solid circle represents the time-average center of mass of grains from the beginning of simulations until the end of interactions (until the center of mass of the ensemble of grains reaches the line crossing the center of the obstacle). The lines are colored in accordance with the initial position of the considered grain with respect to the time-average center of mass, showed in the colorbar on the right. $R$ is the radius of the initial pile, and $r_c$ and $r_0$ are, respectively, the instantaneous and initial positions of the center of mass of the granular material.}
	\label{fig:trajectories}
\end{figure}

Because in this work we carried out numerical simulations in which the position of each particle is known at each time step, we can track all grains, compute their trajectories, and analyze their typical paths. In addition, DEM computes the forces acting on each particle, which means that the instantaneous value of the resultant force acting on individual grains is known, something that cannot be measured in the field or in experiments (at least with the current technologies). We discuss next the trajectories of grains for the three simulated cases and correlate them with the fields of resultant forces. 

Figures \ref{fig:trajectories}a, \ref{fig:trajectories}b and \ref{fig:trajectories}c show the trajectories of individual grains for the bypass, pass over and trapped cases, respectively. For determining which grains are moving, we considered $|\vec{u}_p|$ $>$ 0.1$u_*$ \cite<following>{Wenzel, Assis2} to avoid plotting the trajectories of grains that are almost static. Therefore, their initial positions correspond to the time instant when $|\vec{u}_p|$ $>$ 0.1$u_*$ is first detected for the considered grain, and their ending positions to those when $|\vec{u}_p|$ $\leq$ 0.1$u_*$. In the panels, the small circles represent the starting point of the considered particle, the black crosses represent its final position, the large circle is the mean perimeter of the initial pile, the vertical red line represents the cross section passing by the center of the obstacle, and the pink-solid circle represents the time-average center of mass of grains from the beginning of simulations until the end of interactions presented in the figure (that is, until their center of mass overpass that of the obstacle). The lines are colored in accordance with the initial position of the considered grain with respect to the time-average center of mass, showed in the colorbar on the right, where $R$ is the radius of the initial pile, and $r_c$ and $r_0$ are, respectively, the instantaneous and initial positions of the center of mass of the granular material. We reinforce that the trajectories are plotted in Figure \ref{fig:trajectories} until the instant when the center of mass of the ensemble of grains reaches the line crossing the center of the obstacle. As expected, the trajectories of particles around or over the obstacle is different in each case. Basically: (i) in the bypass case all grains go around the obstacle without touching it, with paths that are significantly curved around the obstacle, and ending close to the recirculation region (from which most of them are afterward entrained downstream); (ii) in the pass over case the grains near the central region of the dune follow paths in the longitudinal direction (almost straight lines), while those in the periphery follow curved paths. The curved paths of peripheral grains occur also for isolated barchans, as shown by \citeA{Alvarez3}; (iii) in the trapped case the grains follow paths approximately circular, migrating toward the recirculation region. Many of them go around the obstacle, while a small part of them go over it (and some of them touch the front part of the obstacle). Most of grains remain in the recirculation region (that is much stronger than in the bypass case), and a small portion is entrained further downstream. We counted the number of grains touching the front part of the obstacle in each case, but this number is only a fraction of the real number since the time step for storing the data is 5000 times that for DEM computations. If we compute proportions related to the trapped case (since it is the larger object), we find that approximately 20 and 0.3\% of grains in the pass over and bypass cases, respectively, touch the front of the obstacle with respect to the trapped case. If this number does not correspond to the total number of grains, it shows at least that in the bypass case the probability of a grain to touch the front of the obstacle is much lower than in the two other cases.

\begin{figure}[!h]
	\begin{center}
		\includegraphics[width=0.7\linewidth]{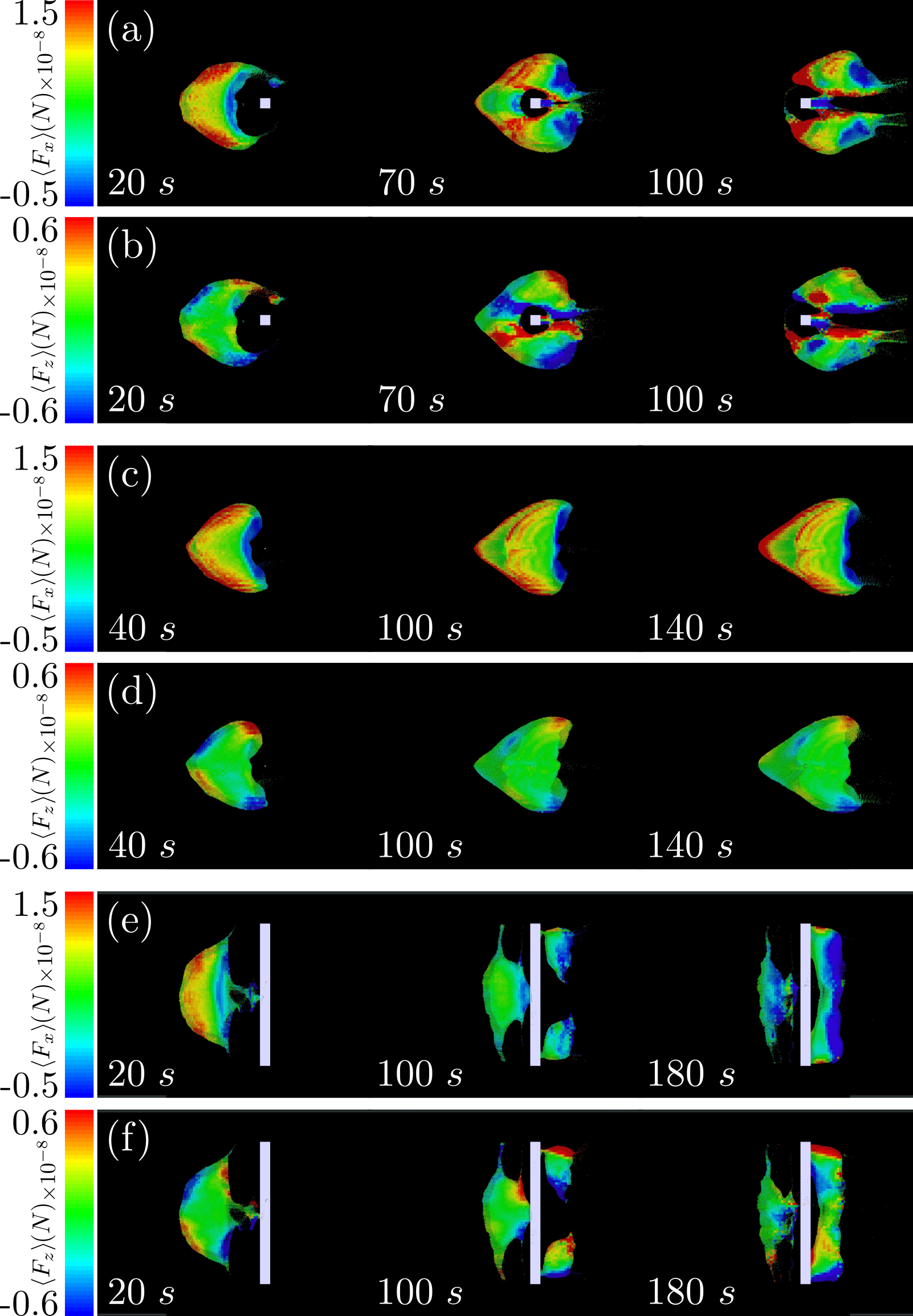}\\
	\end{center}
	\caption{Snapshots showing top view images of a dune colored in accordance with the average level of the resultant force of individual grains. The forces were space averaged in  0.6 mm $\times$ 0.6 mm meshes, and then time averaged over 40 s. (a) and (b) longitudinal $\left< F_x \right>$ and transverse $\left< F_z \right>$ components, respectively, for the bypass case. (c) and (d) $\left< F_x \right>$ and $\left< F_z \right>$, respectively, for the pass over case. (e) and (f) $\left< F_x \right>$ and $\left< F_z \right>$, respectively, for the trapped case. The time instants of the 40 s intervals shown on the bottom left of each snapshot correspond to the mean time instant. The colorbar on the left of each row indicates the force levels.}
	\label{fig:forces}
\end{figure}

In the three cases, we note that the trajectories corroborate the dune behavior, circumventing the obstacle (Figures \ref{fig:morphology}a and \ref{fig:trajectories}a), reaching the obstacle and passing over (or touching) it (Figures \ref{fig:morphology}b and \ref{fig:trajectories}b), or being stretched in the transverse direction (Figures \ref{fig:morphology}c and \ref{fig:trajectories}c). Those trajectories are a direct consequence of the resultant forces acting on each particle, which we investigate next. In order to better visualize the results, we computed time averages of the resultant forces in pre-defined meshes, given the large number of particles in our system (10$^5$ for the initial pile). The resulting space-time averages were computed over 40 s and meshes of 0.6 mm $\times$ 0.6 mm. Figure \ref{fig:forces} shows top view images of a dune colored in accordance with the average level of the resultant force acting on grains (in the considered meshes), in which: Figures \ref{fig:forces}a and \ref{fig:forces}b correspond to the longitudinal $\left< F_x \right>$ and transverse $\left< F_z \right>$ components, respectively, for the bypass case; Figures \ref{fig:forces}c and \ref{fig:forces}d to $\left< F_x \right>$ and $\left< F_z \right>$, respectively, for the pass over case; and Figures \ref{fig:forces}e and \ref{fig:forces}f to $\left< F_x \right>$ and $\left< F_z \right>$, respectively, for the trapped case. The time instants shown on the bottom left of each snapshot correspond to the middle time (instant) within the 40 s intervals, and the colorbar on the left of each row indicates the force levels. According to the colorbar, red and blue correspond to downstream and upstream directions, respectively, for $\left< F_x \right>$, and to negative and positive values with respect to the $z$ coordinate (up and down directions, respectively, in Figures \ref{fig:forces}b, \ref{fig:forces}d and \ref{fig:forces}f) for $\left< F_z \right>$.

In the bypass case (Figures \ref{fig:forces}a and \ref{fig:forces}b), we observe that just before reaching the obstacle ($t$ $=$ 20 s) the grains close to the flanks and upstream the centroid of the barchan experience accelerations that are downstream in the longitudinal direction and outward in the transverse direction, while downstream the centroid an upstream acceleration is observed in the base of the lee face and an inward acceleration occurs in the external part of the horns. When bypassing the obstacle ($t$ $=$ 70 s), grains upstream the obstacle and close to the dune flanks are strongly accelerated downstream and outward, while downstream the obstacle grains are accelerated in the inverse directions: upstream and inward. This is in perfect agreement with the trajectories of particles presented in Figure \ref{fig:trajectories}a. In the pass over case (Figures \ref{fig:forces}c and \ref{fig:forces}d), the distribution of forces just before reaching the obstacle ($t$ $=$ 40 s) is similar to the bypass case, with forces distributed over the dune being somewhat less concentrated. As the barchan passes over the obstacle ($t$ $=$ 100 and 140 s), little change is observed. Finally, for the trapped case (Figures \ref{fig:forces}e and \ref{fig:forces}f), grains are initially ($t$ $=$ 20 s) accelerated downstream and outward, and when circumventing the obstacle ($t$ $=$ 100 and 180 s) are accelerated outward when upstream the obstacle and are accelerated upstream and inward when downstream the obstacle.

\begin{figure}[!h]
	\begin{center}
		\includegraphics[width=0.7\linewidth]{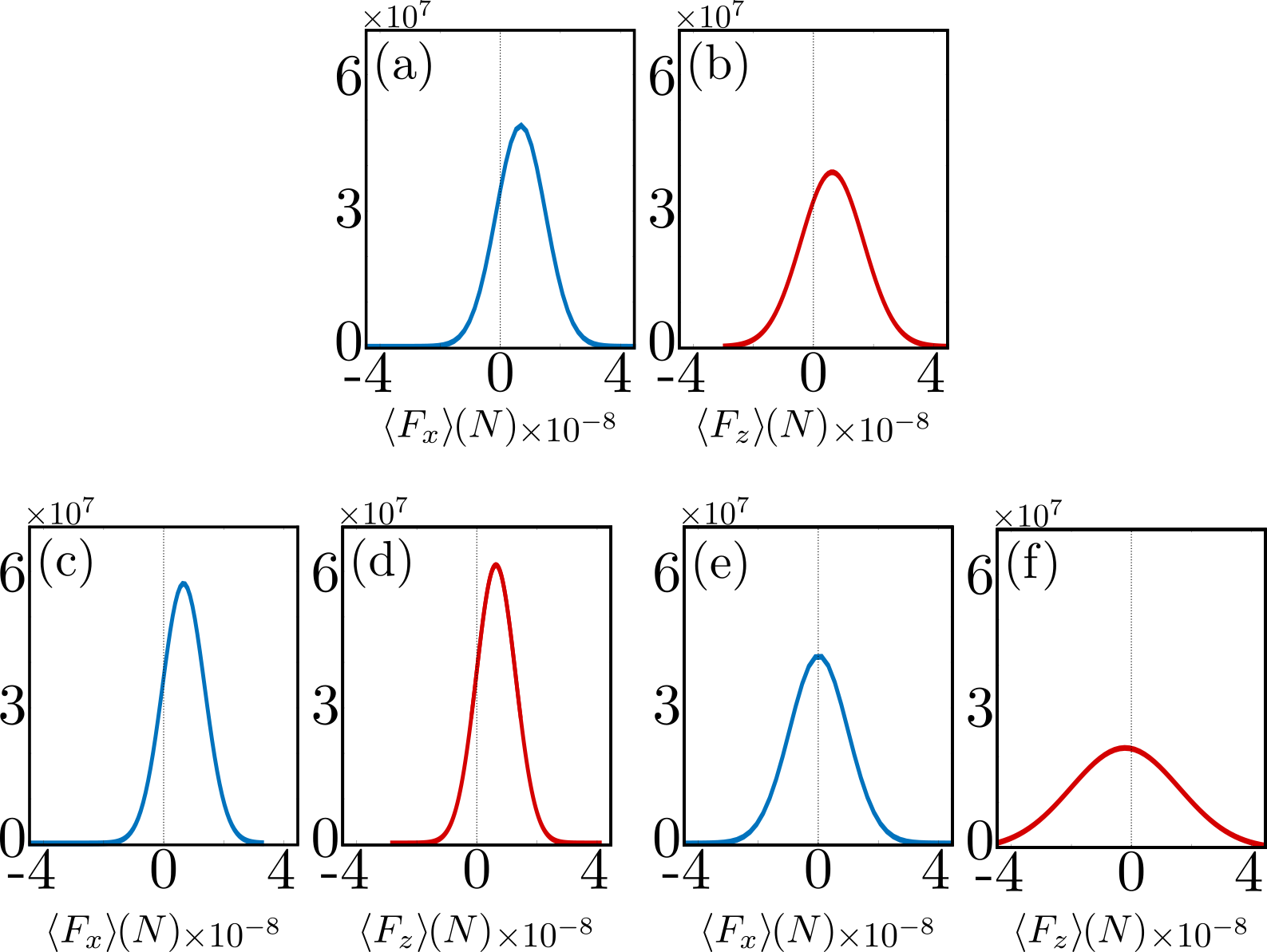}\\
	\end{center}
	\caption{Probability density functions (PDFs) of the longitudinal $\left< F_x \right>$ (in blue) and transverse $\left< F_z \right>$ (in red) components of forces plotted in Figure \ref{fig:forces}. In the Figure, the PDFs in panels (a) and (b) correspond to Figures \ref{fig:forces}a and \ref{fig:forces}b at $t$ $=$ 70s (bypass case), those in panels (c) and (d) to Figures \ref{fig:forces}c and \ref{fig:forces}d at $t$ $=$ 100 s (pass over case), and those in panels (e) and (f) to Figures \ref{fig:forces}e and \ref{fig:forces}f at $t$ $=$ 100 s (trapped case), respectively.}
	\label{fig:force_hist}
\end{figure}

Figure \ref{fig:force_hist} shows the probability density functions (PDFs) of the longitudinal $\left< F_x \right>$ (in blue) and transverse $\left< F_z \right>$ (in red) components of forces plotted in Figure \ref{fig:forces}. Therefore, the PDFs correspond to the spatial distribution of the mean forces for some considerable time intervals: the PDFs in panels (a) and (b) correspond to Figures \ref{fig:forces}a and \ref{fig:forces}b at $t$ $=$ 70s (bypass case), those in panels (c) and (d) to Figures \ref{fig:forces}c and \ref{fig:forces}d at $t$ $=$ 100 s (pass over case), and those in panels (e) and (f) to Figures \ref{fig:forces}e and \ref{fig:forces}f at $t$ $=$ 100 s (trapped case), respectively, where the time instants correspond to the mean time at the considered interval. We note that in the three cases we present the PDFs in the middle of the interaction (obstacle in the middle of the bedform). We observe first that the force distributions are much more concentrated around the most probable value in the pass over case (Figures \ref{fig:force_hist}c and \ref{fig:force_hist}d, with standard deviations of 6.99 $\times$ 10$^{-9}$ and 6.52 $\times$ 10$^{-9}$ N in the $x$ and $z$ directions, respectively), which is consistent with the fact that grains follow paths that are more linear in this case, with less spreading in the transverse direction. We observe the opposite in the trapped case (Figures \ref{fig:force_hist}e and \ref{fig:force_hist}f): the distributions are more spread around the peak, with a large standard deviation (1.80 $\times$ 10$^{-8}$ N) in the case of $\left< F_z \right>$. In addition, in this case the peaks of $\left< F_x \right>$ and $\left< F_z \right>$ are close to zero.  These distributions are consistent with the widening and eventual destruction of the barchan dune: most grains are entrained from the original bedform toward the recirculation region downstream the obstacle (for which grains follow approximately circular paths, as shown in Figure \ref{fig:trajectories}c). Finally, for the bypass case (Figures \ref{fig:force_hist}a and \ref{fig:force_hist}b) the PDFs have a shape that lies within those of the pass over and trapped cases (standard deviations of 8.20 $\times$ 10$^{-9}$ and 1.04 $\times$ 10$^{-8}$ N in the $x$ and $z$ directions, respectively).

\section{Conclusions}

In this paper, we inquired further into the interaction of subaqueous barchans with dune-size obstacles by carrying out numerical simulations. In our computations, the motion of each grain is computed at each time step and the fluid flow is solved down to the scale of grains. The simulations recovered the bypass, pass over and trapped cases reported by \citeA{Assis4}, for which we analyzed the trajectories, velocities, and resultant forces on grains, as well as the fluid flow. We show that the origin of the different behaviors are the strength of flow disturbances with respect to the grains' inertia, this ratio decreasing in the following order: trapped, bypass and pass over. In the bypass case, virtually all grains circumvent the obstacle without touching it, entrained by the flow disturbances (that are responsible for their curved trajectories). In the pass over case, the flow disturbances caused by the obstacle are relatively small, so that grains in the central region of the dune follow straight lines that pass over the obstacle \cite<those in the periphery follow curved lines, as also happens for isolated barchans,>{Alvarez3}. In the trapped case, grains follow closely the fluid flow along pathlines that are approximately circular, some of them remaining trapped in the recirculation region downstream the obstacle while another portion is entrained further downstream by the fluid. The map of resultant forces corroborates the motions of individual grains in all cases. Finally, in the trapped and bypass cases the instantaneous streamlines of the fluid show the presence of a strong vortex between the dune and the obstacle. This vortex results from the interaction between the recirculation region just downstream the dune crest and a horseshoe vortex that exists upstream the obstacle, and it hinders grains from touching or passing over the obstacle. Our findings contribute for the better understanding of the dynamics of dunes in the presence of large obstacles such as hills, craters, and human constructions, situations commonly found both on Earth and Mars. However, due to differences in the relative inertia of grains (and their motion) in aquatic and eolian environments, extrapolations of our results to eolian barchans must be carried out with care.

\section*{Open Research}
\begin{sloppypar}
	Data supporting this work were generated by ourselves and are available in Mendeley Data \cite{Supplemental2} under the CC-BY-4.0 license. The numerical scripts used to post-process the numerical outputs are also available in Mendeley Data \cite{Supplemental2} under the CC-BY-4.0 license.
\end{sloppypar}

\acknowledgments
\begin{sloppypar}
The authors are grateful to the S\~ao Paulo Research Foundation - FAPESP (Grant Nos. 2018/14981-7, 2019/10239-7, 2019/20888-2 and 2022/01758-3) and to the Conselho Nacional de Desenvolvimento Cient\'ifico e Tecnol\'ogico - CNPq (Grant No. 405512/2022-8) for the financial support provided.
\end{sloppypar}

%\nocite{cover_sup, kowalczyk_sup, Pedregosa_sup, rhys_sup}
\bibliography{references}

\end{document}